\documentclass[12pt]{article}
\usepackage{amssymb,amsmath}
\usepackage{latexsym}
\usepackage{graphicx}
\usepackage{cite}
\usepackage[linktocpage=true]{hyperref}

\makeatletter
\@addtoreset{equation}{section}
\makeatother

\def\coeff#1#2{\relax{\textstyle {#1 \over #2}}\displaystyle}

\def\cL{{\cal L}}

\def\cN{{\cal N}}

\begin{document}
\begin{titlepage}

\begin{flushright}

\end{flushright}

\bigskip
\bigskip

\centerline{\Large \bf BTZ Trailing Strings}
\date{\today}
\bigskip
\bigskip
\centerline{{\bf Iosif Bena$^1$ and Alexander Tyukov$^{2,3}$ }}
\bigskip
\centerline{$^1$ Institut de Physique Th\'eorique, Universit\'e Paris Saclay, Orme des Merisiers,}
\centerline{ CEA, CNRS, F-91191 Gif-sur-Yvette, France}
\bigskip
\centerline{$^2$ Dipartimento di Fisica ed Astronomia ``Galileo Galilei," Universit`a di Padova,}
\centerline{Via Marzolo 8, 35131 Padova, Italy} 
\bigskip
\centerline{$^3$ I.N.F.N. Sezione di Padova, Via Marzolo 8, 35131 Padova, Italy}
\bigskip
\bigskip

\begin{abstract}
\noindent

\noindent 

We compute holographically the energy loss of a moving quark in various states of the D1-D5 CFT. In the dual bulk geometries, the quark is the end of a trailing string, and the profile of this string determines the drag force exerted by the medium on the quark. We find no drag force when the CFT state has no momentum, and a nontrivial force for any value of the velocity (even at rest) when the string extends in the supersymmetric D1-D5-P black-hole geometry, or a horizonless microstate geometry thereof. As the length of the throat of the microstate geometry increases, the drag force approaches the thermal BTZ expression, confirming the ability of these microstate geometries to capture typical black-hole physics. We also find that when the D1-D5-P black hole is non-extremal, there is a special value of the velocity at which a moving quark feels no force. We compute this velocity holographically and we compare it to the velocity computed in the CFT.

\end{abstract}

\bigskip

\bigskip

\bigskip

\bigskip

\bigskip

\bigskip

\center{\it Dedicated to the memory of Steve Gubser}

\end{titlepage}

\tableofcontents

\section{Introduction}
One of the features of particles moving in a hot medium is energy loss. For a quark moving in a hot quark-gluon plasma (QGP) this energy loss gives rise to jet quenching: when two back-to-back jets are emitted from near the edge of a QGP blob, the jet that traverses the QGP is significantly attenuated. In the strongly-coupled ${\cal N}=4$ Super-Yang-Mills quark-gluon plasma, the energy loss can also be computed holographically. The quark moving in the hot QGP of this theory is dual to a fundamental string moving in an AdS$_5$-Schwarzschild solution \cite{Herzog:2006gh,Gubser:2006bz}. One end of the string is moving at constant velocity on the boundary, and the other end is trailing near the black hole horizon. The end of string can only have a constant velocity if a constant force is applied, and the energy corresponding to the work exerted by this force travels down the string all the way to the black hole horizon. 

There have been many studies of trailing strings in AdS$_5$ and of the implications of their physics for the quenching that takes place in the real-world QGP. However, the purpose of this paper is to address a more conceptual problem: jet quenching is the only known string theory example of ``friction'' on a particle, and it is important to understand what causes this friction and the associated energy loss\footnote{There exist many examples when the energy of a wave is lost because the wave is absorbed by a black hole, but the jet quenching is the only such energy-loss mechanism for particles.}. In particular, one can ask whether the friction and the associated energy loss only happen because of the presence of a finite-temperature event horizon in the bulk. 

Another important related question is under what circumstances do moving strings acquire an event horizon on their worldsheet. It is well known that this happens when there is a black hole in the bulk \cite{Herzog:2006gh,Gubser:2006bz}, and when the end of the string is accelerated (see for example \cite{Mikhailov:2003er,Fadafan:2008bq,Hubeny:2014kma}). However, it is not known whether a worldsheet horizon appears when there is no spacetime event horizon and no acceleration.

To address these questions we extend the trailing-string calculation to solutions which are asymptotically AdS$_3 \times$ S$^3 \times$ T$^4$. These solutions include the empty (singular) Poincar\'e patch of AdS$_3$, the supersymmetric D1-D5-P black hole, the non-supersymmetric black hole, as well as the supersymmetric smooth horizonless microstate geometries that are dual to pure states of the D1-D5 CFT and that can differ from the supersymmetric black hole arbitrarily close to the horizon. 

Trailing string calculations in supersymmetric backgrounds do not appear at first glance to be very useful, since the only reason the trailing string in AdS$_5$ loses energy is the presence of a horizon whose temperature is that of the dual QGP. However, things are a bit more subtle: AdS$_3$ black holes have both a right- and a left-moving temperature and, in the supersymmetric limit, only one of these temperatures needs be zero. Turning off both temperatures results in a singular solution corresponding to the Poincar\'e patch of AdS$_3$. Hence, one may hope that even in supersymmetric solutions where one of the temperatures is finite there will be nontrivial trailing string profiles and energy loss. 

Our calculations show indeed that in the solution where both the left-moving and the right-moving temperatures are zero, a string coming from the boundary goes straight to the AdS$_3$ horizon and hence there is no energy loss. However, as we will see in Section \ref{sec:BPS}, once a BPS left-moving momentum is turned on, the string starts bending along the direction of the momentum, regardless of the direction in which its end moves on the boundary. Hence the quark corresponding to the end of this string always feels a force in the direction of the momentum in the CFT\footnote{This behavior is different from that quarks in ${\cal N}$=4 QGP, which do not feel a force when at rest.}. This force becomes smaller as the quark moves {\em faster} in the direction of the momentum, and becomes zero in the (unphysical) limit in which the end of the trailing string moves with the speed of light.

There is a very simple ``weakly-coupled'' interpretation for this behavior. The momentum of the D1-D5 system is carried by 1-5 open strings, which move at the speed of light. These strings interact with the heavy quark, and push it in the direction of their momentum. As the quark moves faster and faster, the energy of the photons in its reference frame is reduced by the Doppler effect, and hence the force it feels is smaller. However, this force is always non-zero, except when the speed of the quark approaches the speed of light. 

If the CFT has both a left-moving and a right-moving temperature, the situation is even more interesting. When the two temperatures are equal, one expects the forces exerted by the CFT right-moving and by the left-moving momentum modes to cancel, and hence the quark at rest to feel no force. The bulk trailing string calculation confirms this.

When the temperatures are different, the string ending on a static quark has a nontrivial profile, so the quark feels a net force. However, when the quark moves with a certain constant velocity (given in equation (\ref{noforce-bulk})) the profile of the string in the bulk is straight and the drag force is zero. 

One can try to understand this fact at weak coupling in the boundary theory, by calculating the Doppler-shifted energies of the right-moving and left-moving particles in the rest-frame of the quark, ignoring the interaction between them. This assumption underlies the entropy counting of near-extremal and extremal D1-D5-P black holes \cite{Horowitz:1996fn,Horowitz:1996ay,Danielsson:2001xe}, and allows one to find the velocity of the quark at which there is no drag force  (given in equation (\ref{noforce-CFT})). The weak- and strong-coupling expression for the no-force velocity do not agree, which is not unexpected. What is interesting is that when the quark moves at the bulk no-force velocity, the average frequencies of the CFT right- and left-movers redshifted to the rest-frame of the quark are equal. 

In the second part of this paper we compute the profile of trailing strings in supersymmetric horizonless microstate geometries. These geometries are dual to pure states of the CFT that carry left-moving momentum. Since from the CFT perspective one expects a moving quark to lose or gain energy by interacting with this momentum, one expects that the strings moving in the bulk would also bend and to lead to an energy loss/gain similar to that of the black hole.
However, the geometries in which the string moves do not have a horizon, and hence the presence of a trailing string would contradict the standard lore that energy loss comes because the presence of a horizon. 

Our calculation indicates that a trailing string exists whenever the momentum is finite, despite the absence of an event horizon. Hence the quark moving against the momentum in a CFT state dual to horizonless geometry will also lose energy. It is only when the string moves in a geometry that has neither left- nor right-moving momentum (such as the original Lunin-Mathur supertube) that the string profile is straight and momentum loss is absent. 

One can also compare the momentum loss in a horizonless geometry to the momentum loss in the BTZ solution. We find that as the throat of the microstate geometries becomes larger and larger, the horizonless trailing sting result approaches closer and closer the BTZ one. Furthermore, the correction to the BTZ result depends on the orientation of the string on the $S^3$, and for the longest possible throats it is suppressed as $1/N_1 N_5$  compared to the leading term. 

The fact that the energy loss in the microstate geometries with long throats only differs from the energy loss in the black holes by terms that are  $1/N_1 N_5$ suppressed has been encountered before, both when studying certain limits of Wightman functions \cite{Raju:2018xue}, and when evaluating correlators of two light and two heavy operators in these geometries \cite{Bena:2019azk}. It indicates that these geometries reproduce with high accuracy the features one expects of the bulk configurations dual to the typical microstates black hole.

Furthermore, despite the absence of a horizon in spacetime, the strings moving in microstate geometries have a  horizon on the worldsheet, exactly like strings moving in the AdS$_5$-Schwarzschild solution. This worldsheet horizon is responsible for the Brownian motion felt by the quark\cite{deBoer:2008gu, Giecold:2009cg} and this Brownian motion is indistinguishable from that felt in the black hole solution. This again confirms the fact that the microstate geometries with long throats behave as typical black hole microstates are supposed to do.

In section \ref{BHsec} we compute trailing string profiles in black hole geometries and evaluate the energy loss/gain of a moving quark. In section \ref{microsec} we repeat these calculations for certain smooth horizonless microstate geometries and compare the results to the black-hole results.

\section{String in D1-D5-P black hole backgrounds}
\label{BHsec}

In this section we consider the motion of a classical F1 string in the background of extremal and non-extremal D1-D5-P black holes. The string does not extend along the sphere or along the compact direction, so it will feel just the BTZ part of this geometry.

\subsection{The extremal BTZ geometry}
\label{sec:BPS}

The metric of the supersymmetric D1-D5-P system is given by (see for example \cite{Bena:2013pda, David:2002wn})
\begin{equation}
	ds_{10}^2 ~=~ (Z_1 Z_5)^{-1/2}(-dt^2+dy^2) + (Z_1 Z_5)^{-1/2}(Z_p-1)(dt+dy)^2 +(Z_1 Z_5)^{1/2}dx_{i}dx^{i} + Z_1^{1/2} Z_5^{-1/2}dx_{a}dx^{a}\,,   
\end{equation}
where $Z_{1,5,p} = 1+{r_{1,5,p}^2}/{r^2}$, $i=1,2,3,4$ and $a=6,7,8,9$. This configuration corresponds to D1 branes stretching along $x^5\equiv y$ and D5 branes along $x^{5,6,7,8,9}$ with the left-moving momentum along the common circle $y$. 

The near-horizon limit corresponds essentially to dropping ``1'' in the $Z_1$ and $Z_5$ harmonic functions, and the metric becomes that of a four-torus times the six-dimensional extremal BTZ $\times $ S$^3$ metric:
\begin{equation}    
	ds_{6}^2 ~=~ \frac{r^2}{L^2}(-dt^2 + dy^2) + \frac{R^2}{L^2}(dt + dy)^2  + \frac{L^2}{r^2}dr^2 + L^2 d\Omega_3^2 \,,
	\label{bpsmetric}
\end{equation}
where $L\equiv\sqrt{r_1 r_5}$ and $R\equiv r_p$. 

We consider a string whose endpoint on the boundary of AdS$_3$ moves at constant velocity, $v$, for a long time, which means that the string profile remains self-similar and does not change with time. We will parametrize the worldsheet by coordinates $\tau = t$ and $\sigma = r$. Then the analysis of the string motion reduces to finding the profile function $\zeta(r)$: 
\begin{equation}
	y(t,r) = v t + \zeta(r)\,,
	\label{embedding}
\end{equation}
which we normalize such that $\zeta(r\to\infty)=0$. 

The induced worldsheet metric $g_{\alpha\beta}=G_{\mu\nu} \partial_{\alpha} X^{\mu} \partial_{\beta} X^{\nu}$ is given by 
\begin{eqnarray}
	ds_{\rm ws}^2 ~=~ \frac{1}{L^2} \Bigg( \left( R^2(1+v)^2 - (1-v^2)r^2 \right) d\tau^2 &+& \left( \frac{L^4}{r^2}+ \zeta'(r)^2(r^2+R^2) \right) d\sigma^2 \nonumber \\ 
	 &+& 2\zeta'(r) \left(R^2+v (r^2+R^2) \right) d\tau d\sigma \Bigg)\,,
\end{eqnarray}
and the Nambu-Goto action is
\begin{equation}
	S ~=~ -T_0 \int{dr \,  \sqrt{1 - v^2 + \frac{r^4}{L^4} \left(\zeta'(r)\right)^2 - \frac{R^2}{r^2}\left(1+v\right)^2}} 
		~=~  T_0 \int{dr \, \cL} \,.
	\label{NG}
\end{equation}
%
Since the string Lagrangian, $\cL$, is invariant under shifts $\zeta(r)\to\zeta(r)+\mbox{const}$, there is a conserved canonical momentum
\begin{equation}
	\pi_{\zeta} ~=~ \frac{\partial \cL}{\partial \zeta'} ~=~ -\frac{r^4 \zeta '(r)}{L^4 \sqrt{1 - v^2 + \frac{r^4}{L^4} \left(\zeta'(r)\right)^2 - \frac{R^2}{r^2}\left(1+v\right)^2} }\,.
\end{equation}
Using this we can solve for $\zeta'(r)$:
\begin{equation}
	\zeta'(r) = \frac{\pi_{\zeta}L^4}{r^3}\sqrt{\frac{r^2 (1-v^2) - R^2(1+v)^2}{r^4- L^4 \pi_{\zeta}^2}}\,,
	\label{der}
\end{equation}  

The above expression looks very similar to that for trailing strings in the AdS$_5$-Schwarzschild geometry \cite{Gubser:2006bz, Herzog:2006gh} and the analysis can be done in a similar fashion, remembering only that the extremal BTZ horizon is at $r=0$.

For the string profile to be well defined one has to fix the value of $\pi_\zeta$ such that the square root in (\ref{der}) is real-valued for all positive values of $r$. The numerator under the square root changes sign at $r_{*}=\gamma R(1+v)$, $\gamma = \frac{1}{\sqrt{1-v^2}}$, and therefore the denominator has to flip sign at $r_{*}$ too. The physical meaning of $r_{*}$ can be easily understood by looking at the worldsheet metric: $r_{*}$ is a location of the ``horizon'' on the string worldsheet, where the components $g_{\tau\tau}$ and $g^{\sigma\sigma}$ flip the sign. This fixes the first integral   
\begin{equation}
	\pi_{\zeta} ~=~ \pm \frac{R^2(1+v)}{L^2(1-v)}\,,  
\label{pi}	
\end{equation}
and gives a simple equation for the string profile
\begin{equation}
	\zeta'(r) ~=~ \frac{\pi_{\zeta} L^4 (1-v^2)}{r^3\sqrt{r^2(1-v^2)+R^2(1+v)^2}}\,.
\end{equation}
The square root on the right hand side is now well-defined for all $r>0$, and the previous equation can be integrated analytically to obtain $\zeta(r)$. 

The drag force can be found from the $y$-component of the spacetime energy-momentum density of the string:
\begin{equation}
	\frac{dp_y}{dt} ~=~ \Pi_{y}^{\sigma}\,,
	\label{dragforce}
\end{equation}
where $p_y$ is the quark momentum and 
\begin{equation}
	\Pi_{\mu}^{\sigma} ~=~ -T_0\, G_{\mu\nu}\, \frac{g_{\tau\sigma}\partial_{\tau} X^{\nu} - g_{\tau\tau}\partial_{\sigma} X^{\nu}}{\sqrt{-g}}\,.
\end{equation}      
This gives the rate of the momentum change: 
\begin{equation}
	\frac{dp_y}{dt} ~=~ \Pi_{y}^{\sigma} ~=~  -T_0\frac{r^4 \zeta '(r)}{L^4 \sqrt{-g}} ~=~  - T_0 \pi_{\zeta}\,.
	\label{force}
\end{equation}
One can also compute the rate of the energy flow along the string:
\begin{equation}
	\Pi^{\sigma}_t ~=~ T_0\frac{r^4 v \zeta '(r)}{L^4 \sqrt{-g}} ~=~ T_0 \pi_{\zeta} v \,. 
	\label{energyflow}
\end{equation}

Both the sign of the force acting on string and the sign of the energy flux depend on the sign of $\pi_{\zeta}$ and, much like in the AdS$_5$ trailing string calculation, it is important to choose the sign of $\pi_{\zeta}$  which gives the correct physics.

If $\pi_{\zeta}$ is positive the string bends along the direction of momentum, and the force on a string is always negative:
\begin{equation}
	\frac{dp_y}{dt} ~=~ - T_0 \frac{R^2(1+v)}{L^2(1-v)} \,.
	\label{force1}
\end{equation}
This means that the right-moving string experiences a drag force caused by the left-moving momentum excitations, and the string loses its momentum. The energy then flows from the boundary into the bulk. 

Eventually the string will lose all its momentum and reverse its motion. The force still will be acting in the same direction, but now instead of dragging it backwards the string will be accelerated forward, along the momentum direction. As seen from (\ref{energyflow}) the direction of the energy flux will also change, and the energy will flow from the bulk to the boundary, accelerating the left-moving quark. In the extreme unphysical case when $v=-1$, corresponding to a quark moving at the speed of light to the left, the force on a quark vanishes, and the profile becomes a straight line stretched from the boundary into the bulk.  

\begin{figure}[t]
\centering
\includegraphics[width=7cm]{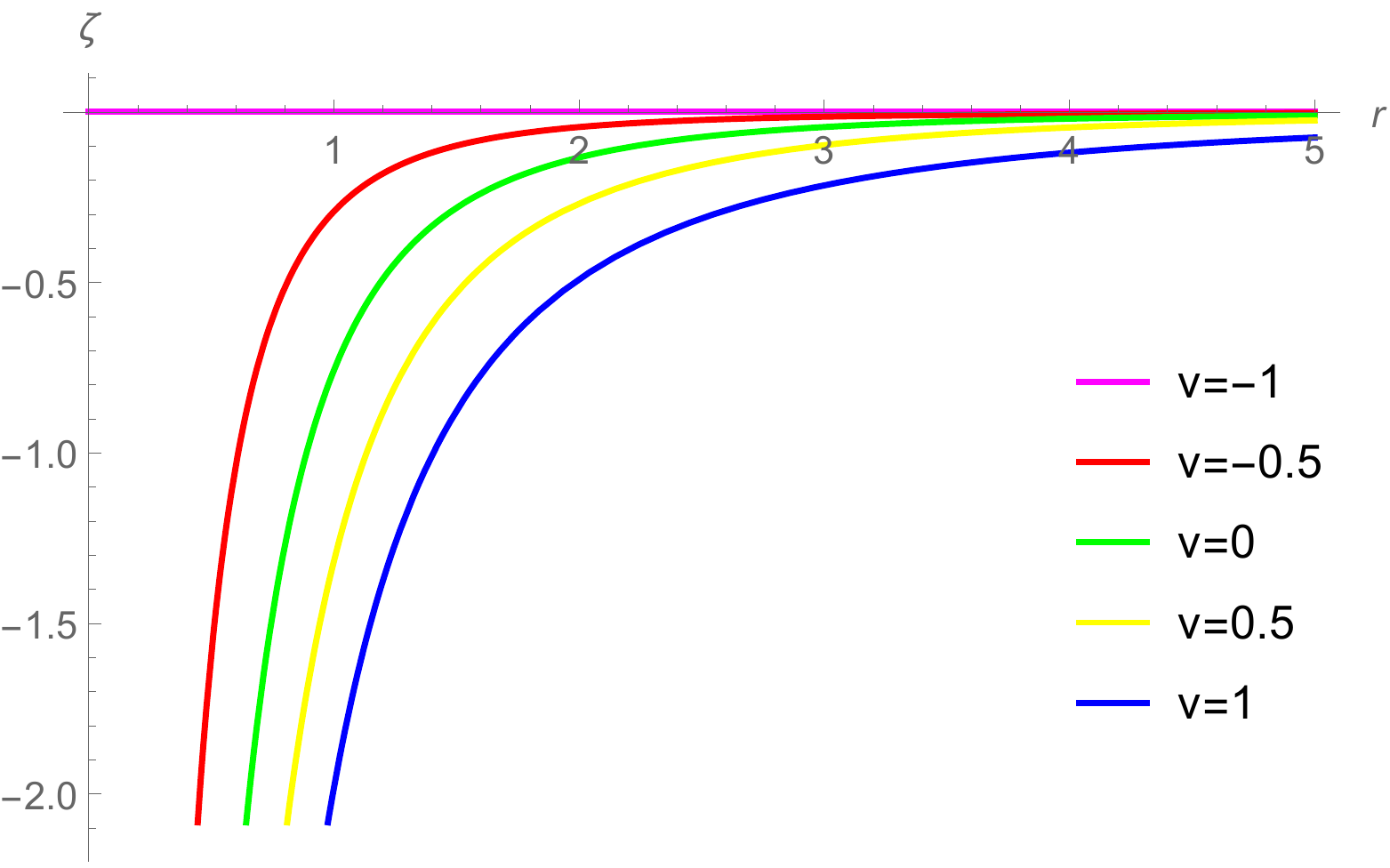} 
\caption{The string profile, $\zeta(r)$, for different values of $v$. The string moves against the momentum (up), for $v>0$ and along the momentum (down) for $v<0$. }
\label{profile}
\end{figure}

In contrast, in the  branch with negative $\pi_{\zeta}$ the string moving against the momentum would be bent forward, and thus the quark corresponding to the end of the string would accelerate. This is clearly unphysical, and hence the physical branch is the one where $\pi_{\zeta}$ is always positive. The plot of the string profiles for different velocities is shown in the Figure \ref{profile}. 


\subsection{The non-extremal BTZ geometry}
\label{sec:nonBPS}

The results of the previous section can be generalized to the non-extremal D1-D5-P black hole. This allows us to study the combined effect of the non-extremality and the momentum on the string motion. The near-horizon limit of the metric is $\text{BTZ}\times S^3 \times T^4$ (see \cite{David:2002wn})
\begin{equation}    
	ds_{10}^2 ~=~ \frac{r^2}{L^2}(-dt^2 + dy^2) + \frac{R^2}{L^2}(\cosh\lambda\, dt + \sinh\lambda\, dy)^2  + \frac{L^2}{r^2-R^2}dr^2 + L^2 d\Omega_3^2 + ds_{T^4}^2\,.
	\label{nonbpsmetric}
\end{equation}
Using the embedding ansatz (\ref{embedding}) it is easy to obtain the induced metric on the string worldsheet:  
\begin{eqnarray}
	& &ds_{\rm ws}^2 = \frac{1}{L^2} \Bigg( \left( R^2 (\cosh\lambda+ v \sinh\lambda)^2 - \left(1-v^2\right) r^2 \right) d\tau^2 \\
	&+&\!\!\!\! \left(\frac{L^4}{r^2-R^2} + \zeta '(r)^2 \left(r^2+R^2 \sinh ^2\lambda\right) \right) d\sigma^2  
	 + 2\zeta '(r) \left(R^2 \sinh\lambda(\cosh\lambda+v \sinh\lambda)+v r^2\right) d\tau d\sigma \Bigg)\,\nonumber
\end{eqnarray}
The string Lagrangian is: 
\begin{equation}
	\cL ~=~ -\frac{1}{L^2}\sqrt{\frac{L^4 \left(\left(1-v^2\right) r^2-R^2 (\cosh\lambda+v \sinh\lambda)^2\right)-r^2 \left(r^2-R^2\right)^2 \zeta '(r)^2}{r^2-R^2}} \,. 
\end{equation}
The derivative of the string profile function becomes:
\begin{equation}
	\zeta'(r) ~=~ \frac{L^4 \pi_{\zeta} \sqrt{r^2\left(1-v^2\right) - R^2 ( \cosh \lambda + v \sinh \lambda)^2}}{r \left(r^2-R^2\right) \sqrt{ r^4-r^2 R^2 - L^4 \pi_{\zeta}^2}}\,,
\end{equation}   
where  $\pi_{\zeta}$ is the canonical momentum and, as before, the critical value $r_{*}$ coincides with the location of the string worldsheet horizon: $r_{*}=\gamma R (\cosh\lambda+v  \sinh\lambda)$. Again, to make the function $\zeta(r)$ real-valued one has to fix the momentum:
\begin{equation}
	\pi_{\zeta} ~=~ \frac{R^2}{ L^2}\frac{(v\cosh\lambda + \sinh\lambda)(\cosh\lambda + v\sinh\lambda)}{\left(1-v^2\right)}\,.
\end{equation}
The equation for the profile function now becomes:
\begin{equation}
	\zeta'(r) ~=~ \frac{L^2 R^2 \left(\left(1+v^2\right) \sinh 2 \lambda +2 v \cosh 2 \lambda \right)}{r(r^2-R^2)\sqrt{2\left(1-v^2\right) \left(2 r^2-R^2\right)+2 R^2 \left(\left(1+v^2\right) \cosh 2 \lambda +2 v \sinh 2 \lambda \right)}}\,,
\end{equation}
which is well-defined for all $r>R$. 

Then the drag force (\ref{dragforce}) is 
\begin{equation}
	\frac{dp_y}{dt} ~=~ - T_0 \frac{R^2}{ L^2}\frac{(v\cosh\lambda + \sinh\lambda)(\cosh\lambda + v\sinh\lambda)}{\left(1-v^2\right)}\,.
	\label{force2}
\end{equation}

The metric (\ref{nonbpsmetric}) has two important limits. The first is the $\lambda\to \infty$ limit, which corresponds to the BPS black hole. In this limit the drag force (\ref{force2}) reproduces the expression (\ref{force1}) if one rescales $R$ as $R\to R\, e^{-\lambda} /2$. The second limit, $\lambda\to 0$, gives the AdS$_3$-Schwarzschild black hole with the horizon radius $R$. The drag force then becomes:
\begin{equation}
	\frac{dp_y}{dt} ~=~ - T_0 \frac{R^2}{ L^2}\frac{v}{1-v^2}\,,
\label{}
\end{equation}
which is in agreement with the drag force in an AdS$_{d+1}$-Schwarzschild black hole computed in \cite{Herzog:2006gh} for $d=2$. 

\subsection{The no-force velocity}
\label{sec:no-force}

An interesting feature of equation \eqref{force2} is that the drag force becomes zero, and the string profile in the bulk is a straight line from the boundary to the horizon, when the quark velocity is
\begin{equation}
v = v_{\rm no-force}^{\rm bulk}\equiv -\tanh  \lambda\,.
\label{noforce-bulk}
\end{equation}
The purpose of this section is to understand  whether one can compute this velocity in the boundary theory at weak coupling.

To do this we should remember that the entropy of the non-extremal D1-D5-P black hole can be computed at weak coupling by assuming the microstates of this black hole to be given by two non-interacting one-dimensional systems of left-moving and right-moving D1-D5 open strings \cite{Horowitz:1996fn, Horowitz:1996ay}, each of which has its own temperature, $T_L$ and $T_R$. The relation between these temperatures and the left- and right momenta, $P_{L,R}$, is \cite{Mandal:2000rp}
\begin{equation}
	T_{L,R} = \frac{1}{\pi R_y} \sqrt{\frac{P_{L,R}}{Q_1 Q_5}}\,,
\end{equation}
and one can express the ratio of the temperatures in terms of the parameter, $\lambda$, of the black hole as 
\begin{equation}
	{T_{R} \over T_{L}} =  e^{-2 \lambda} \,.
\end{equation}

The average energy of the massless bosonic excitations in the gas equals its temperature, up to some numerical constant. Indeed, in any dimension, $D$, the total number of particles goes like $T^D$, whereas the total energy goes like $T^{D+1}$. Hence, the average frequencies of these modes are
\begin{equation}
	\omega_{L,R} = {\frac{1}{R_y} }\sqrt{\frac{P_{L,R}}{Q_1 Q_5}}\,,
\end{equation}
and the average numbers of left- and right-moving particles are
\begin{equation}
	n_{L,R} = \sqrt{P_{L,R} Q_1 Q_5}\,.
\end{equation}

One can then compute the Doppler-shifted average energies of left- and right-movers in the rest-frame of the quark:
\begin{equation}
	\omega'_{R} = \omega_R \sqrt{{\frac{1-v}{1+v}}}\,, \quad \omega'_{L} = \omega_L \sqrt{{\frac{1+v}{1-v}}}\,,  
\end{equation}
where the quark is moving to the right with velocity $v$. Interestingly, the ratio of the Doppler-shifted frequencies of the left- and right-movers:
\begin{equation}
	{\omega'_{R} \over \omega'_{L}} = \left({\frac{1-v}{1+v}}\right) {\omega_{R} \over \omega_{L}} =  {\frac{1-v}{1+v}} e^{-2 \lambda}\,,
\end{equation}
becomes one when the velocity matches the no-force velocity \eqref{noforce-bulk} calculated in the bulk using the trailing string, $v=-\tanh \lambda$.

However, the pressures exerted on the quark, $\Pi_{R,L}$, are equal to these Doppler-shifted frequencies  multiplied by the average numbers of particles, and hence
\begin{equation}
	{\Pi_{R} \over \Pi_{L}} = {n_{R} ~\omega'_{R} \over n_{L} ~\omega'_{L}}  =  {P_R \over P_L} {\frac{1-v}{1+v}} =
	{\frac{1-v}{1+v}} e^{-4 \lambda}\,.
\end{equation}
Thus, at weak coupling the pressures are equal and the quark feels no force when 
\begin{equation}
v = v_{\rm no-force}^{\rm boundary}\equiv -\tanh 2 \lambda\,.
\label{noforce-CFT}
\end{equation}

The no-force velocity computed holographically at strong coupling \eqref{noforce-bulk} and the one computed in the boundary theory at weak coupling \eqref{noforce-CFT} do not agree. Since this is not a quantity protected in any way, this mismatch is not surprising. In particular it implies that the weak-coupling picture of the right-moving CFT states, as a gas of right-moving massless particles with a thermal distribution of energies, does not work at strong coupling.

\section{Strings in microstate geometries}
\label{microsec}

In this section we consider the motion of a string in certain families of horizonless microstate geometries \cite{Bena:2017upb}, that have the same charges and angular momenta as a D1-D5-P black hole. These geometries, which have a long 
BTZ-like throat that ends with a smooth cap instead of an event horizon were constructed using superstratum technology \cite{Bena:2015bea}. In the dual D1-D5 CFT they are dual to certain family of pure momentum-carrying states \cite{Bena:2016ypk,Bena:2017xbt}. 

\subsection{The family of metrics}

The metric can be written in as fibration of $S^3$ over a $(2+1)$-dimensional base: 
\begin{eqnarray}
 ds_6^2 &=& \sqrt{Q_1 Q_5} \,  \frac{\Lambda}{F_2(r)} \bigg[ \frac{F_2(r) \, dr^2}{r^2 + a^2}  \,-\, \frac{2\,F_1(r)}{a^2 (2 a^2 + b^2)^2 \;\! R_y^2 }\bigg(dv + \frac{a^2\,(a^4 + (2 a^2 +b^2) r^2)}{F_1(r)} du \bigg)^2 \cr
& & \qquad \quad \qquad \qquad~ +\, \frac{2 \, a^2 \,r^2 \,(r^2 + a^2)\, F_2(r)}{F_1(r) \, R_y^2} \, du^2 \bigg] \label{sixmet} \\
&& ~+\sqrt{Q_1 Q_5} \, \bigg[  \Lambda \, d\theta^2  \,+\,  
\frac{1}{\Lambda} \sin^2 \theta \, \Big(d\varphi_1 -  \frac{a^2}{(2a^2 +b^2)}\frac{\sqrt{2}}{R_y} (du + dv) \Big)^2  \cr
&& \qquad\qquad\quad~~ + \, \frac{F_2(r)}{\Lambda} \cos^2 \theta \, \Big(d\varphi_2 + \, \frac{1}{(2a^2 +b^2)\, F_2(r)}\frac{\sqrt{2}}{R_y}  \left[ a^2 (du - dv) -  b^2 \, F_0 (r) dv \right] \Big)^2  \bigg] \,,
\nonumber
\end{eqnarray}
where the functions, $F_i(r)$, are
\begin{equation}
\begin{aligned}
F_0(r) ~\equiv~ & 1 - \frac{r^{2n}}{(r^2 +a^2)^{n}}  \,, \qquad F_1(r) ~\equiv~ a^6 - b^2 \, (2 a^2 + b^2) \,r^2 \, F_0(r) \,, \\
F_2(r) ~\equiv~ &  1  - \frac{a^2\, b^2}{(2 a^2 + b^2) } \,\frac{ r^{2n}}{(r^2 +a^2)^{n+1}}   \, ,
\end{aligned}
  \label{Fdefs}
\end{equation}
and the warp factor, $\Lambda$, is
\begin{equation}
\Lambda ~\equiv~ \sqrt{ 1 - \frac{a^2\,b^2}{(2 a^2 +b^2)} \, \frac{r^{2n}}{(r^2 +a^2)^{n+1}} \, \sin^2 \theta  } \,.
  \label{Lambdadef1}
\end{equation}
The coordinates $u$ and $v$ are the standard light-cone coordinates:
\begin{equation}
	u ~=~ \frac{1}{\sqrt{2}}(t-y)\,, \qquad v ~=~ \frac{1}{\sqrt{2}}(t+y)\,,
\end{equation}
where $y$ is a coordinate on $S^1$ with the period
\begin{equation}
	y ~\equiv~ y + 2\pi R_y\,.
\end{equation}

The parameters, $a$ and $b$, are related to the conserved five-dimensional angular momenta and $y$-momentum: 
\begin{equation}
J ~\equiv~ J_L ~=~ J_R ~=~  \coeff{1}{2}\, \cN \,  a^2  \,, \qquad \qquad N_P   ~=~  \coeff{1}{2}\, \cN \,n \, b^2\,,
\end{equation}
where $\cN = N_1 N_5 R_y^2/ (Q_1 Q_5) $, $N_1$, $N_5$ are the numbers of D1 and D5 branes and $Q_1$ and $Q_5$ are their supergravity charges. They also have to satisfy the condition coming from requiring the solution to be regular:
\begin{equation} 
\frac{Q_1Q_5}{R_y^2} ~=~ a^2 + \coeff{1}{2}\, b^2 \,.
\end{equation} 

The metric closely approximates the BTZ geometry but caps off smoothly with no horizon. The depth of the throat is controlled by the parameter $a$, and in the $a\to 0$ limit the metric becomes the one of the extremal BTZ times $S^3$:
\begin{equation}
 ds_6^2  ~=~ \sqrt{Q_1 Q_5} \,  \bigg(  \bigg[ \, \frac{d\rho^2}{\rho^2}  - \rho^2 \, dt^2 +  \rho^2 \, dy^2 + \frac{n}{R_y^2} \, (dy+dt)^2 \bigg] 
~+~
\big[  d\theta^2  +  \sin^2 \theta \, d\varphi_1^2 + \cos^2 \theta  \, d\varphi_2^2 \, \big] \bigg)  \label{BTZmet} \,,
\end{equation}
where $\rho \equiv (Q_1 Q_5)^{-\frac{1}{2}} \,r$.

\subsection{String motion}

As a first approximation we will ignore the interaction of the string with the NS-NS B-field. This can be justified because this B-field, which is non-zero in microstate geometries, is proportional to oscillating Fourier modes, and therefore its RMS value is zero. Thus we are still going to use the Nambu-Goto action to describe the string motion.        
	
We will consider only the $n=1$ metric, because the analysis of the string Lagrangian gets more involved for higher $n$. However, we believe that $n=1$ solution captures the significant features of string motion in microstate geometries.

Since the metric on the base depends on the position of the string on the $S^3$, we first choose to place the string at the South Pole, $\theta=\pi/2$. The warp factor, $\Lambda$, maximally deviates from $1$ at $\theta=\pi/2$, and thus one should expect the strongest influence on the string motion. We are still going to use the embedding given by (\ref{embedding}), although it is possible to consider a combined motion on the base and on the $S^3$. 

As before, one can compute the induced metric on the worldsheet:  
\begin{equation}
\begin{split}
	ds_{\rm ws}^2 ~=~ \frac{1}{\sqrt{2} R \sqrt{\left(2 a^2+b^2\right) \left(a^2+r^2\right)^2-a^2 b^2 r^2}} \Bigg(  (1+v) r^2 \left(b^2 (1+v)-2 (1-v)(a^2+r^2)  \right)  d\tau^2 \\
	+ \left(R^2 \left(2 a^2+b^2 - \frac{a^2 b^2 r^2}{\left(a^2+r^2\right)^2}\right) + r^2 \left(b^2+2(a^2+r^2) \right) \zeta'(r)^2 \right) d\sigma^2  \\
	+ 2 r^2\zeta '(r)  \left(b^2 (1+v)+ 2 v\left(a^2+r^2\right)\right) d\tau d\sigma \Bigg)\,,
\end{split}	
\end{equation}
and the corresponding Nambu-Goto Lagrangian:
\begin{equation}
	\cL ~=~ -\sqrt{\frac{(1+v) r^2 \left(b^2 (1+v)-2 (1-v) \left(a^2+r^2\right)\right)}{2 \left(a^2+r^2\right)^2}-\frac{2 r^4 \left(a^2+r^2\right)^2 \zeta'(r)^2}{R^2 \left(\left(2 a^2+b^2\right) \left(a^2+r^2\right)^2-a^2 b^2 r^2\right)}}\,.
\end{equation}

One can also find the derivative of the string profile in terms of the conserved canonical momentum: 
\begin{equation}
	\zeta '(r) ~=~ \frac{\pi_{\zeta} R_y^2 \left(\left(2 a^2+b^2\right) \left(a^2+r^2\right)^2-a^2 b^2 r^2\right) \sqrt{(1+v) \left(2 (1-v) \left(a^2+r^2\right)-b^2 (1+v)\right)}}{2 r \left(a^2+r^2\right)^2 \sqrt{2 r^4 \left(a^2+r^2\right)^2-\pi_{\zeta}^2 R_y^2 \left(\left(2 a^2+b^2\right) \left(a^2+r^2\right)^2-a^2 b^2 r^2\right)}}\,.
\label{zetaprime}	
\end{equation}
The analysis of the above expression shows that critical value of $r$ for the square root in the numerator is 
\begin{equation}
	r_{\ast}^2 ~=~ \frac{b^2 (1+v)}{2 (1-v)}-a^2\,,
\end{equation}
which determines the worldsheet horizon of the string. Since $r_{\ast}^2$ has to be non-negative, this places a constraint on the parameters $a$ and $b$ of the metric. Assuming $v\geq 0$ it is easy to verify that $r_{\ast}^2 $ is always non-negative if 
\begin{equation}
	b \geq \sqrt{2}\, a\,. 
	\label{constraint}
\end{equation}
It is interesting to note that this bound lies inside the so-called ``black-hole regime'' of microstates \cite{Bena:2016ypk,Bena:2017xbt}:
\begin{equation}
	\frac{b^2}{a^2}>\frac{k}{n+\sqrt{(k-m+n)(m+n)}}\,,
\end{equation}
which in an $(1,0,1)$ superstratum metric is: $b>\sqrt{\sqrt{2}-1}\,a$. Therefore, one should expect similar physics to that of a supersymmetric black hole. 
  
The expression under the square root in the denominator will change sign at $r_{\ast}$ if one picks the value of the canonical momentum:
\begin{equation}
	\pi_{\zeta} ~=~ \frac{b (1+v) \left(b^2 (1+v)-2 a^2 (1-v)\right)}{ R_y (1-v)\sqrt{8 a^4 (1-v)^2+8 a^2 b^2 v  (1+v)+2 b^4 (1+v)^2}}\,.
\end{equation}
It is possible to show that if one substitutes the above value of $\pi_{\zeta}$ back in (\ref{zetaprime}), the resulting expression will be real-valued for all values of $r$, provided that (\ref{constraint}) is satisfied. 

\begin{figure}[t]
\centering
\includegraphics[width=7cm]{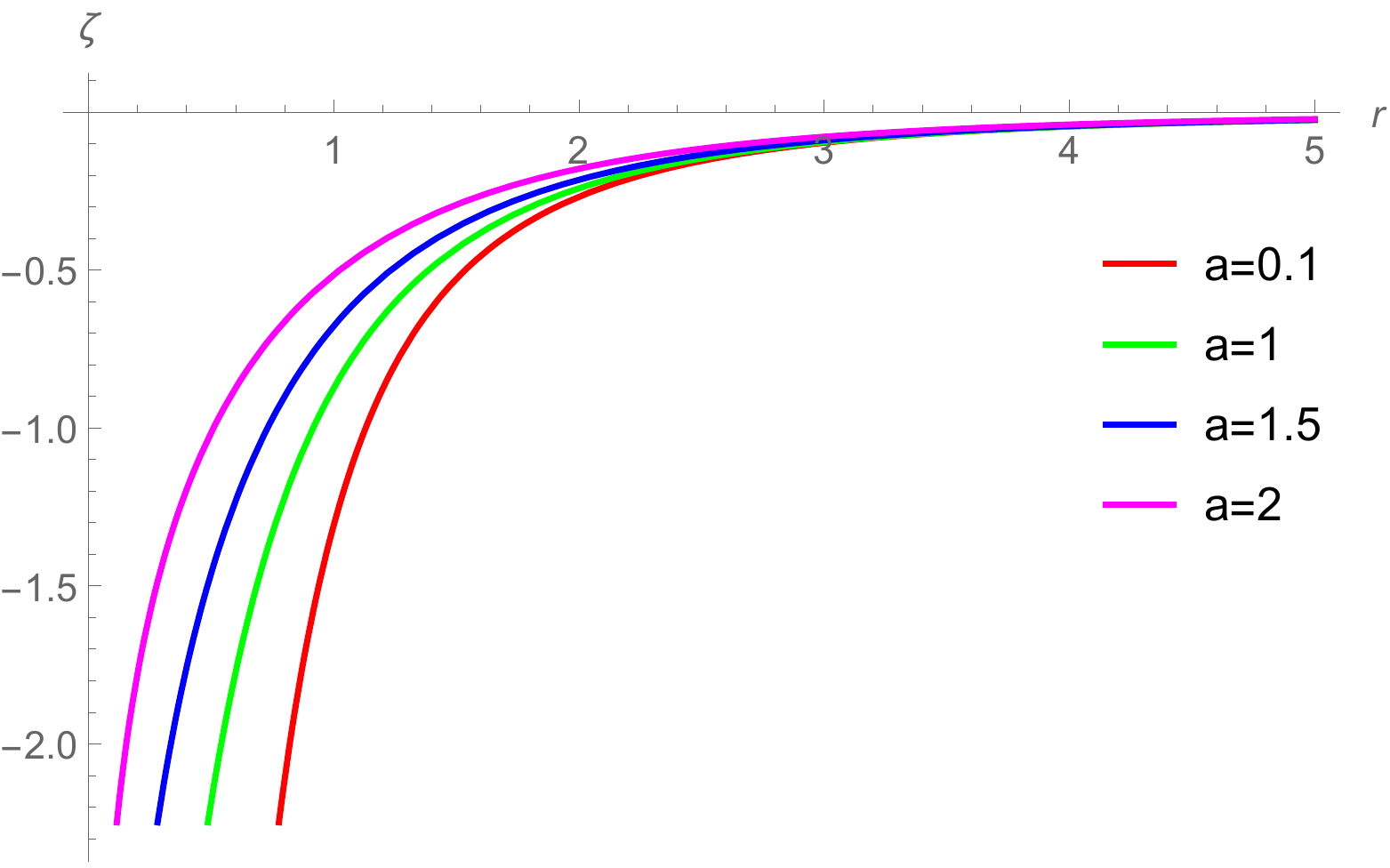}
\caption{The string profile $\zeta(r)$ in the microstate geometry for $v=0.5$ and different values of $a$.}
\label{msprf}
\end{figure}

It is easy to see from (\ref{zetaprime}) that the string profile is logarithmically divergent around $r=0$, and thus the string will asymptotically approach $r=0$. In Fig. \ref{msprf} we plot the result of the numerical integration of the string profile for different values of $a$ which controls the angular momentum of the solution. 

The drag force exerted on a string is
\begin{equation}
	\frac{dp_y}{dt} ~=~ - T_0 \pi_{\zeta} ~=~ - T_0 \frac{b (1+v) \left(b^2 (1+v) - 2 a^2 (1-v)\right)}{ R_y (1-v)\sqrt{8 a^4 (1-v)^2+8 a^2 b^2 v  (1+v)+2 b^4 (1+v)^2}}\,.
\end{equation}
As before there is no momentum transfer for $v=-1$, and the string is straight. 

As we mentioned earlier the microstate geometry approximates the black hole arbitrary close, and the $a\to 0$ limit of the metric is the extremal BTZ times $S^3$. The parameters of black hole and microstate solution at $a\to 0$ are related: $b =\sqrt{2} R$, and $R_y=L^2/R$. It is therefore interesting to compare the microstate drag force to the black hole one. The power expansion of the former for small values of $a$ gives:  
\begin{equation}
	\frac{dp_y}{dt}  ~\approx~ - T_0 \left(\frac{b (1+v)}{\sqrt{2} R_y (1-v)}-\frac{\sqrt{2} a^2 }{b R_y (1-v)}+O\left(a^4\right) \right)\,,
\label{force3}	
\end{equation}
which can be considered as a finite correction to (\ref{force1}) coming from replacing the horizon with a smooth cap. This correction term is suppressed by the factor of $a^2/b^2$ compared to the black-hole result which for the "deep, scaling geometries" with $a\ll b$ can be approximated by $J/N_1 N_5$. For the longest possible throat $J=1$, and the correction is $1/N_1 N_5$ smaller compared to the leading term.     

Unlike the BTZ trailing string, the microstate trailing string end does not asymptotically approach the horizon simply because there is none. Instead it approaches the ``supertube locus" $(r=0, \theta=\pi/2)$. 

It is also instructive to consider a string at $\theta=0$, which is the North Pole of the sphere. The woldsheet geometry simplifies greatly as the warp factor, $\Lambda$, becomes $1$. Following the very same steps as before one can get an equation for the string profile: 
\begin{equation}
	\zeta '(r) ~=~ \frac{\pi_{\zeta}  R_y^2 \left(2 a^2+b^2\right) \sqrt{(1+v) \left(2(1-v)(r^2+a^2)-b^2(1+v)\right)}}{2 \left(a^2+r^2\right)^{\frac32} \sqrt{2 \left(a^2+r^2\right)^2-\pi_{\zeta} ^2 R_y^2 \left(2 a^2+b^2\right) }}\,,
\end{equation}
which leads to the same critical value $r_{\ast}$. Now the constant of integration has to be chosen as
\begin{equation}
	\pi_{\zeta} ~=~ \frac{b^2 (1+v)}{\sqrt{2}R_y (1-v)  \sqrt{2 a^2+b^2}}\,,
\label{piconst}
\end{equation}
and the string profile becomes:
\begin{equation}
	\zeta '(r) ~=~ \frac{1}{2} b^2 R_y \left(\frac{1+v}{a^2+r^2}\right)^{3/2} \sqrt{\frac{2 a^2+b^2}{b^2(1+v)+2(1-v)(r^2+a^2)}}\,.
\end{equation}      

This profile can be integrated, and one can show that the string reaches the bottom of the geometry, the $(r=0,\theta=0)$ point. 
However, the slope of the string in the $(y,r)$-plane is not zero at this point and hence the string does not asymptote to this location (as it happens for strings at $\theta=\pi/2$). Rather this point is at a finite distance on the string worldsheet from the worldsheet horizon.  Since there is no object at this location, the string cannot end there. Hence, there should be a way to continue the string beyond the $(r=0,\theta=0)$ point. 

To see this, we remember that the construction of superstrata \cite{Bena:2015bea} makes use of the spheroidal or two-centered coordinates in which the $\mathbb{R}^4$ base of the solution is parameterized as 
\begin{equation}
	x_1+i x_2 = \sqrt{r^2+a^2} \, \sin \theta \, e^{i \phi}\,, \quad x_3+i x_4 = r \,  \cos \theta \,  e^{i\psi}\,.
\end{equation}
In this coordinates $r=0$ corresponds to a disk of radius $a$ parametrized by $\theta$ and $\phi$. The ``supertube locus" is at the edge of this disk $(r=0,\theta=\pi/2)$.   

Once the string reaches $r=0$ it is natural to assume that the string continues along the $\theta$-direction. Since the embedding (\ref{embedding}) degenerates at $r=0$, the embedding of the remaining part of the string will be  \begin{equation}
	y(t,\theta) = v t + \xi(\theta)\,.
	\label{embedding2}
\end{equation} 
 
As before we can compute the Lagrangian for the string:  
\begin{equation}
	\mathcal{L} ~=~ -\sqrt{\frac{1}{2} (1+v) \left(2 a^2 (1-v)-b^2 (1+v)\right) \cos ^2 \theta + \frac{2 a^4 \cos ^4 \theta \,  \xi '(\theta )^2}{R_y^2 \left(2 a^2+b^2\right)}}\,, 
\end{equation}
from which one gets the new string profile: 
\begin{equation}
	\xi'(\theta) ~=~ \frac{ \pi_{\xi}  R_y^2 \left(2 a^2+b^2\right) \sqrt{(1+v) \left(b^2 (1+v)\right)-2 a^2 (1-v)}}{2 a^2 \cos\theta \sqrt{\pi_{\xi} ^2 R_y^2 \left(2 a^2+b^2\right)-2 a^4 \cos ^4 \theta }}\,.
\end{equation}

Now the integral of motion $\pi_{\xi}$ should be fixed by matching this derivative with $\zeta'(r)$ at the point $(r=0,\theta=0)$. If we assume that $\pi_{\xi}=\pi_{\zeta}$ and use the expression (\ref{piconst}) we obtain 
\begin{equation}
	\xi'(\theta) ~=~ \frac{b^2 (1+v)^{\frac32} R_y \sqrt{\left(2 a^2+b^2\right)\left( b^2(1+v)-2a^2(1-v) \right)}}{2 a^2 \cos \theta \sqrt{b^4 (1+v)^2-4 a^4 (1-v)^2 \cos ^4 \theta}}\,,
\end{equation}
with the matching condition $\xi'(\theta=0)=a\, \zeta'(r=0)$. This suggests the following identification of worldsheet coordinate: $\sigma=a\,\theta$. The square root in the denominator is well defined for all values of $\theta\in[0,\pi/2]$ provided $v\geq 0$ and $a$,$b$ satisfy the constraint (\ref{constraint}). The string profile is logarithmically divergent at $\theta=\pi/2$. Interestingly enough, the coefficient of the leading divergent term of $\xi(\theta)$ coincides with the one of $\xi(r)$ at $(r=0,\theta=\pi/2)$ point which can be seen from the expansion of (\ref{zetaprime}) around $r=0$. This strongly supports the fact that the end of the trailing string will asymptotically approach the ``supertube locus" $(r=0,\theta=\pi/2)$ irrespectively of where the string is placed on the sphere.

The frictional force on the string is again determined by the value of $\pi_{\zeta}$:
\begin{equation}
	\frac{dp_y}{dt}  ~=~ - T_0 \pi_{\zeta} ~=~ T_0 \frac{b^2 (1+v)}{\sqrt{2}R_y (1-v)  \sqrt{2 a^2+b^2}}\,,   
\end{equation}
which in the long-throat (small $a$) limit becomes:         
\begin{equation}
	\frac{dp_y}{dt}  ~=~ - T_0  \left( \frac{b (1+v)}{ \sqrt{2} R_y (1-v)}-\frac{a^2 (1+v)}{ \sqrt{2} b R_y (1-v)}+O\left(a^4\right) \right) \,.
\end{equation}

Again, the leading term in the expansion reproduces the frictional force felt by strings moving in an extremal black hole. However, the $a^2$-correction slightly differs from the one in (\ref{force3}) and hence depends on the location of the trailing string inside the $S^3$.     

\bigskip

\noindent{\bf Acknowledgments}\\
We would like to thank Edmond Iancu, David Mateos and Nick Warner for interesting discussions.  The work of I.B, has been supported by the ANR grant Black-dS-String ANR-16-CE31-0004-01, the John Templeton Foundation grant 61169 and the ERC Grant 787320 - QBH Structure. The work of A.T. has been supported in part by the MIUR-PRIN contract 2017CC72MK\_003.
A.T. would like to thank the IPhT, CEA-Saclay for hospitality during various stages of this collaboration.

\newpage



\begin{thebibliography}{99}


\bibitem{Gubser:2006bz}
  S.~S.~Gubser,
  ``Drag force in AdS/CFT,''
  Phys.\ Rev.\ D {\bf 74} (2006) 126005
  doi:10.1103/PhysRevD.74.126005
  [hep-th/0605182].

\bibitem{Herzog:2006gh} 
  C.~P.~Herzog, A.~Karch, P.~Kovtun, C.~Kozcaz and L.~G.~Yaffe,
  ``Energy loss of a heavy quark moving through N=4 supersymmetric Yang-Mills plasma,''
  JHEP {\bf 0607}, 013 (2006)
  doi:10.1088/1126-6708/2006/07/013
  [hep-th/0605158].


\bibitem{Hubeny:2014kma}
  V.~E.~Hubeny and G.~W.~Semenoff,
  ``String worldsheet for accelerating quark,''
  JHEP {\bf 1510} (2015) 071
  doi:10.1007/JHEP10(2015)071
  [arXiv:1410.1171 [hep-th]].

\bibitem{Mikhailov:2003er} 
  A.~Mikhailov,
  ``Nonlinear waves in AdS / CFT correspondence,''
  hep-th/0305196.


\bibitem{Fadafan:2008bq} 
  K.~Bitaghsir Fadafan, H.~Liu, K.~Rajagopal and U.~A.~Wiedemann,
  ``Stirring Strongly Coupled Plasma,''
  Eur.\ Phys.\ J.\ C {\bf 61}, 553 (2009)
  doi:10.1140/epjc/s10052-009-0885-6
  [arXiv:0809.2869 [hep-ph]].

\bibitem{Horowitz:1996fn} 
  G.~T.~Horowitz and A.~Strominger,
  ``Counting states of near extremal black holes,''
  Phys.\ Rev.\ Lett.\  {\bf 77}, 2368 (1996)
  doi:10.1103/PhysRevLett.77.2368
  [hep-th/9602051].


\bibitem{Horowitz:1996ay} 
  G.~T.~Horowitz, J.~M.~Maldacena and A.~Strominger,
  ``Nonextremal black hole microstates and U duality,''
  Phys.\ Lett.\ B {\bf 383}, 151 (1996)
  doi:10.1016/0370-2693(96)00738-1
  [hep-th/9603109].

\bibitem{Danielsson:2001xe} 
  U.~H.~Danielsson, A.~Guijosa and M.~Kruczenski,
  ``Brane anti-brane systems at finite temperature and the entropy of black branes,''
  JHEP {\bf 0109}, 011 (2001)
  doi:10.1088/1126-6708/2001/09/011
  [hep-th/0106201].



\bibitem{Raju:2018xue} 
  S.~Raju and P.~Shrivastava,
  ``Critique of the fuzzball program,''
  Phys.\ Rev.\ D {\bf 99}, no. 6, 066009 (2019)
  doi:10.1103/PhysRevD.99.066009
  [arXiv:1804.10616 [hep-th]].

\bibitem{Bena:2019azk}
  I.~Bena, P.~Heidmann, R.~Monten and N.~P.~Warner,
  ``Thermal Decay without Information Loss in Horizonless Microstate Geometries,''
  SciPost Phys.\  {\bf 7} (2019) 063
  doi:10.21468/SciPostPhys.7.5.063
  [arXiv:1905.05194 [hep-th]].

	
\bibitem{deBoer:2008gu} 
  J.~de Boer, V.~E.~Hubeny, M.~Rangamani and M.~Shigemori,
  ``Brownian motion in AdS/CFT,''
  JHEP {\bf 0907}, 094 (2009)
  doi:10.1088/1126-6708/2009/07/094
  [arXiv:0812.5112 [hep-th]].
  
\bibitem{Giecold:2009cg} 
  G.~C.~Giecold, E.~Iancu and A.~H.~Mueller,
  ``Stochastic trailing string and Langevin dynamics from AdS/CFT,''
  JHEP {\bf 0907}, 033 (2009)
  doi:10.1088/1126-6708/2009/07/033
  [arXiv:0903.1840 [hep-th]].



\bibitem{Bena:2013pda} 
  I.~Bena, S.~El-Showk and B.~Vercnocke,
  ``Black Holes in String Theory,''
  Springer Proc.\ Phys.\  {\bf 144}, 59 (2013).
  doi:10.1007/978-3-319-00215-6-2

\bibitem{David:2002wn} 
  J.~R.~David, G.~Mandal and S.~R.~Wadia,
  ``Microscopic formulation of black holes in string theory,''
  Phys.\ Rept.\  {\bf 369}, 549 (2002)
  doi:10.1016/S0370-1573(02)00271-5
  [hep-th/0203048].

\bibitem{Mandal:2000rp}
  G.~Mandal,
  ``A Review of the D1 / D5 system and five-dimensional black hole from supergravity and brane viewpoint,''
  hep-th/0002184.

\bibitem{Bena:2017upb} 
  I.~Bena, D.~Turton, R.~Walker and N.~P.~Warner,
  ``Integrability and Black-Hole Microstate Geometries,''
  JHEP {\bf 1711}, 021 (2017)
  doi:10.1007/JHEP11(2017)021
  [arXiv:1709.01107 [hep-th]].
     

\bibitem{Bena:2015bea} 
  I.~Bena, S.~Giusto, R.~Russo, M.~Shigemori and N.~P.~Warner,
  ``Habemus Superstratum! A constructive proof of the existence of superstrata,''
  JHEP {\bf 1505}, 110 (2015)
  doi:10.1007/JHEP05(2015)110
  [arXiv:1503.01463 [hep-th]].
	
\bibitem{Bena:2016ypk}
  I.~Bena, S.~Giusto, E.~J.~Martinec, R.~Russo, M.~Shigemori, D.~Turton and N.~P.~Warner,
  ``Smooth horizonless geometries deep inside the black-hole regime,''
  Phys.\ Rev.\ Lett.\  {\bf 117} (2016) no.20,  201601
  doi:10.1103/PhysRevLett.117.201601
  [arXiv:1607.03908 [hep-th]].
		
\bibitem{Bena:2017xbt}
  I.~Bena, S.~Giusto, E.~J.~Martinec, R.~Russo, M.~Shigemori, D.~Turton and N.~P.~Warner,
  ``Asymptotically-flat supergravity solutions deep inside the black-hole regime,''
  JHEP {\bf 1802} (2018) 014
  doi:10.1007/JHEP02(2018)014
  [arXiv:1711.10474 [hep-th]].
	
	
	
	
\end{thebibliography}
\end{document}